\documentclass[prl,aps,twocolumn,superscriptaddress]{revtex4-1}
\usepackage{graphicx,color}
\usepackage{amsthm}
\usepackage{amsfonts}
\usepackage{algorithmic}
\usepackage{enumerate}
\usepackage{latexsym}
\usepackage{amsmath}
\usepackage{amssymb}
\usepackage[colorlinks,linkcolor=blue,anchorcolor=blue,citecolor=blue,urlcolor=blue]{hyperref}
\def\avg#1{\left\langle#1\right\rangle}

\def\avg#1{\langle#1\rangle}
\def\Re{\rm{Re}}
\def\Im{\rm{Im}}
\def\be{\begin{equation}} \def\ee{\end{equation}}
\def\bea{\begin{eqnarray}} \def\eea{\end{eqnarray}}
\def\PRB{Phys. Rev. B}
\def\PRA{Phys. Rev. A}
\def\PRL{Phys. Rev. Lett.}
\def\nn{\nonumber}
\def\pp{\parallel}

\emergencystretch=\maxdimen
\begin{document}
\title{Doping-driven Antiferromagnetic Insulator -
Superconductor Transition: a Quantum Monte-Carlo Study}
\author{Tianxing Ma}
\affiliation{Department of Physics, Beijing Normal University,
Beijing 100875, China}
\affiliation{Department of Physics, University of California,
San Diego, California 92093, USA}
\author{Da Wang}
\affiliation{National Laboratory of Solid State Microstructures $\&$
School of Physics, Nanjing University, Nanjing 210093, China}
\author{Congjun Wu}
\affiliation{Department of Physics, University of California,
San Diego, California 92093, USA}

\begin{abstract}
How superconductivity emerges in the vicinity of an antiferromagnetic
insulating state is a long-standing issue of strong correlation
physics.
We study the transition from an antiferromagnetic insulator to a
superconductor by hole-doping based on a bilayer generalization of
a Hubbard-like model.
The projector quantum Monte-Carlo simulations are employed, which are sign-problem-free both at and away from half-filling.
An anisotropic Ising antiferromagnetic Mott insulating phase
occurs at half-filling, which is weakened by hole-doping.
Below a critical doping value,
antiferromagnetism coexists with the singlet superconductivity,
which is a pairing across each rung with an extended $s$-wave symmetry.
As further increasing doping, the antiferromagnetic order vanishes,
leaving only a superconducting phase.
These results provide important information on how superconductivity
appears upon doping the parent Mott-insulating state.
\end{abstract}
\maketitle

\noindent
\emph{Introduction}~
The study on strongly correlated electron systems is a central topic of condensed matter physics for exploring novel states of matter.
In the vicinity of the antiferromagnetic (AF) insulating
phase, unconventional superconducting (SC) states appear by doping,
or, applying pressure to systems of heavy-fermion materials \cite{Steglich1979}, high $T_c$ cuprates \cite{Bednorz1986},
iron pnictides \cite{Kamihara2008}, and organic superconductors
\cite{Jerome1980}.
In the past several decades, the doped Mott insulators and the consequential competitions among antiferromagnetism, superconductivity, and charge
orderings have been extensively studied with significant
efforts from various different perspectives \cite{Zhang1997,Sachdev2003,*Kivelson2003,*Lee2006,
weng2011,chen2014,Demler2004,Fradkin2015,Zheng1155,Huang1161,Jiang1424}.

How superconductivity arises by doping Mott-insulators is an outstanding problem of condensed matter physics.
{
Due to its non-perturbative nature, sufficiently accurate
numerical methods are essential to resolve small energy differences
among competing orders} \cite{Scalapino2012,Zheng1155,Huang1161,Jiang1424}.
Nevertheless, exact diagonalizations are limited to small system sizes due
to the exponential growth of the many-body Hilbert space \cite{Dagotto1994}.
The density-matrix-renormalization group \cite{White1992}
and tensor-network methods \cite{Schollwoeck2011} have been
successfully applied to two-dimensional (2D) spin models \cite{Stoudenmire2012}
and quasi-one-dimensional fermionic ladder systems \cite{Liu2012,White2015}.
However, their applications to 2D fermionic systems are just beginning \cite{Jiang1424,Corboz2014}.
The results of the variational Monte Carlo method depend on the input
trial wavefunctions \cite{Edegger2007}.
The auxiliary field quantum Monte-Carlo (QMC) method \cite{Hirsch1983,Blankenbecler1981} is unbiased, but
it suffers from the notorious sign-problem when doping away from
half-filling \cite{Troyer2005}.
Once the sign-problem occurs, the numeric errors grow exponentially
as enlarging the system size and lowering the temperature,
which usually plagues simulations, corresponding to a regime of maximal numerical difficulty in computational science for decades
\cite{Zheng1155}.

Recently, a progress has appeared to employ the auxiliary field QMC method to study a spin-fermion model \cite{Abanov2000},
which describes the low energy hot-spot fermionic excitations
and yields the $d$-wave like pairing symmetry \cite{Berg2012}.
This model is designed to be sign-problem free based on the previously
proved Kramers-invariant decomposition
by one of the authors and Zhang \cite{Wu2005}.
In such a decomposition, the Hubbard-Stratonovich (HS) transformation to fermion interactions
is formulated in a Kramers invariant way, {\it i.e.}, the fermion
matrix in any HS field configuration
satisfying the Kramers symmetry.
Its determinant, working as the statistical weight, is a product of complex-conjugate pairs, and thus positive-definite.
Developments along this line mainly follow the hot-spot
dominated pairing mechanism \cite{Schattner2016,Li2017,Wang2017}.
However, these models begin with a metallic normal state
far away from the Mott physics.
For microscopic models such as the Hubbard-like ones exhibiting
Mott-physics at half-filling, QMC simulations contribute
significantly to the study of pairing mechanisms \cite{White1989,Hirsch1989},
nevertheless, they often suffer from the notorious sign-problem upon doping.
It is desired to simulate the emergence of superconductivity by doping
Mott insulators through QMC simulations in a sign-problem free way.

In this article, we investigate the competition between antiferromagnetism
and superconductivity by doping the parent 2D
Mott-insulators.
We employ the Scalapino-Zhang-Hanke (SZH) model by generalizing
it to a bilayer version.
It is a Hubbard-like model augmented by charge and spin-exchange interactions {
across each  rung consisting of two sites}.
In a wide range of interaction parameters, it satisfies the criterion of
the Kramers invariant decomposition for QMC simulations \cite{Wu2005},
hence, is sign-problem free at arbitrary electron fillings.
This enables the possibility to study the transition from the AF insulating state to the SC state {
in a numerically exact manner,
that is, any accuracy can be achieved within a polynomial time.}
At half-filling, the ground state is either an AF insulator
in the case with the Ising anisotropy, or, a rung-singlet Mott phase
with the SU(2) invariance.
Upon hole doping, the AF ordering is weakened and finally
suppressed when the doping level $x>x_c\approx 0.11$.
Meanwhile, the extended $s$-wave SC order grows up away from
half-filling and coexists with the AF order at $0<x<x_c$.

\begin{figure}
\includegraphics[width=0.3\textwidth]{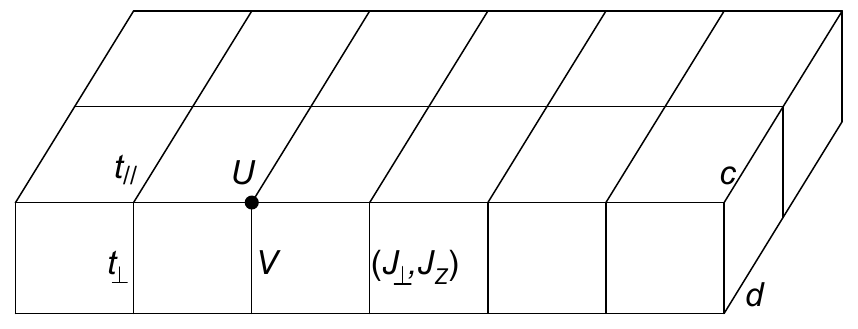}
\caption{\label{Fig:sketch}
The generalized SZH model is defined on a bilayer square lattice.
Its parameters include the intra- and inter-layer hoppings of
$t_\parallel$ and $t_\perp$, respectively, the onsite Hubbard interaction $U$,
the interactions between two sites along each rung with $V$ in the
charge channel, $J_\perp$ and $J_z$ of the AF super-exchanges
along the  $x(y)$ and $z$-directions, respectively.
}
\end{figure}

\noindent
\emph{Model and QMC Simulations.}
The SZH model \cite{Scalapino1998}, originally
defined for a two-leg ladder, is an extended
Hubbard model for studying competing orders
in strongly correlated systems.
We further generalize it to a bilayer square lattice as sketched
in Fig.~\ref{Fig:sketch}.
The Hamiltonian reads,
\begin{eqnarray}
H&=&-t_{||}\sum_{{\langle ij\rangle}\sigma}(c^\dagger_{i\sigma}c_{j\sigma}+d^\dagger_{i\sigma}d_{j\sigma}+H.c.) \notag \\
&-&t_\perp\sum_{i\sigma}(c^\dagger_{i\sigma}d_{i\sigma}+H.c.) -
\mu\sum_{i\sigma}(c^\dagger_{i\sigma}c_{i\sigma}+ d^\dagger_{i \sigma}d_{i\sigma})
\notag \\
&+&\frac{J_\perp}{2}\sum_{i}(S^{+}_{ic}S^{-}_{id}+h.c.)+J_z\sum_{i}S^{z}_{ic}S^{z}_{id} \notag \\
&+&U\sum_{i}[(n_{i\uparrow c}-\frac{1}{2})(n_{i\downarrow c}-\frac{1}{2})
+(n_{i\uparrow d}-\frac{1}{2})(n_{i \downarrow d}-\frac{1}{2})] \notag \\
&+&V\sum_{i}(n_{ic}-1)(n_{id}-1)
\label{eq:SZHham},
\end{eqnarray}
where the electron annihilation operators in the upper and lower layers are denoted
as $c$ and $d$, respectively.
The Hamiltonian Eq. \ref{eq:SZHham} consists of the intra- and inter-layer nearest-neighboring hopping terms of $t_\parallel$ and $t_\perp$, respectively,
and $t_\parallel$ is the energy unit set as 1 throughout this article.
The interaction terms include the onsite Hubbard interaction $U$,
and the interactions between two sites along each
vertical rung: $V$ is the charge channel interaction; $J_\perp$ and $J_z$
are the transverse and longitudinal spin exchanges, respectively.
The in-plane AF correlation is intermediated through the 2nd
order perturbation theory.
For the isotropic case with $J_\perp = J_z$, the system enters the rung
singlet phase at half-filling when the super-exchange interaction
across each rung is larger than the in-plane one which is perturbatively
small.
We first consider the case with the Ising anisotropy by setting $J_z > J_\perp$
which stabilizes the AF long-range-order along the $z$-direction
at half-filling. The AF parent insulating state is then doped for
achieving the SC phase.

{
We will use Eq. \ref{eq:SZHham} for studying a hard-core strong correlation problem in 2D that how superconductivity emerges by doping Mott insulators.
The major advantage is that such a model will be later shown to be QMC sign-problem free in a large parameter region, hence, it can be studied in a numerically exact way.
Furthermore, it can be mapped to a monolayer two-orbital model \cite{bilayer_model_2011,bilayer_model_2013} with the two orbitals
equivalent to the upper and lower layers, respectively.
Multi-orbital models have been widely studied in strongly
correlated systems such as iron-based superconductors \cite{iron_rmp}.
It can also be mapped to a spin-$\frac{3}{2}$ fermionic
Hubbard model \cite{Wu2003,Wu2005,Capponi2004,Wu2006}
in a compact way by defining}
$\psi_i=[c_{i\uparrow},c_{i\downarrow},d_{i\uparrow},d_{i\downarrow}]^t$,
\begin{eqnarray}
H&=&-t_{||}\sum_{\langle ij \rangle} \left( \psi_i^\dag \psi_j + h.c.\right)
-t_\perp\sum_i \psi_i^\dag\Gamma^5\psi_i - \mu \sum_i n_i\nn \\
&-&\sum_i \frac{g_c}{2} \left(n_i-2\right)^2
-\sum_{i, a=1\sim 5} \frac{g_a}{2} \left(n_i^a \right)^2,
\label{eq:ham32}
\end{eqnarray}
where
\bea
n_i=\psi_i^\dag \psi_i, \ \ \, n_i^a= \frac{1}{2}\psi_i^\dag \Gamma^a \psi_i,
\eea
and the five $\Gamma$-matrices are the rank-2 Clifford algebra,
satisfying $\{ \Gamma^a, \Gamma^b\}=2\delta_{ab}$ with $1\le a <b \le 5$,
as defined in the Supplemental Material (S.M.) I following the convention
in Ref.~[\onlinecite{Wu2003}].
The interaction parameters in the two different representations
of Eq. \ref{eq:SZHham} and Eq. \ref{eq:ham32} are related by
\bea
&&4g_c=\frac{J_\perp}{2}+\frac{J_z}{4}-U-3V, \notag\\
&&g_{1,5}=\frac{J_\perp}{2}+\frac{J_z}{4}-U+V,  \notag \\
&&g_{2,3}=\frac{J_\perp}{2}-\frac{J_z}{4}+U-V,  \notag\\
&&g_4=-\frac{J_\perp}{2}+\frac{3J_z}{4}+U-V.
\eea
In Eq. \ref{eq:ham32}, all the interaction terms are expressed in
Kramers invariant operators $n_i$ and $n_i^a$, which satisfy
\bea
\mathcal{T} n_i \mathcal{T}^{-1}=n_i, \ \ \,
\mathcal{T} n^a_i \mathcal{T}^{-1}=n^a_i,
\eea
and
the Kramers transformation is defined as
$\mathcal{T}=\Gamma^1\Gamma^3\mathcal{C}$ ($\mathcal{C}$
means complex conjugate).
$\mathcal{T}$ is the usual time-reversal
transformation followed by switching the upper and lower layers.

When all the coupling constants $g_c$ and $g_a$~ $(1\le a \le 5)$
are non-negative, the HS decomposition can be
performed in a Kramers invariant way, such that the auxiliary field
QMC is free of the sign problem \cite{Wu2003,Wu2005,Capponi2004,Wu2006}.
The discrete HS decomposition for the
4-fermion interaction is performed in an exact way as shown in S. M. II.
Roughly speaking, $g_c$ favors the charge-density-wave (CDW) order, and
$g_{1,5}$ favors the rung current, or, bond-wave order, respectively
\cite{Capponi2004},
while $g_{2,3,4}$ favors the AF order.
For simplicity, we set $g_c=g_1=g_5=0$ for studying the
antiferromagnetism-superconductivity transition.
In practice, we have chosen
$ U=\frac{5}{2},  J_\perp=1, J_z=8, V=t_\perp=0$,
corresponding to $g_4=8g_2=8g_3=8$.
Our QMC simulations employ the projector scheme working at
zero temperature with the projection time $\beta=4L$ and the discrete imaginary
time slice $\Delta\tau=0.1$.
The results show convergences with respect to $\beta$ and $\Delta\tau$
as shown in the S. M. III, respectively
\footnote{
The choice of $\beta=4L$ means that as $L$ increases, the projection time $\beta$ increases accordingly, ensuring the ground state can be achieved for each $L$ as the gap is roughly proportional to $1/L$.
}.
These simulations are performed on 20 cores for each group of
parameters with 500 warm-up steps and more than 1000 steps
of measurements.

\begin{figure}
\includegraphics[width=0.45\textwidth]{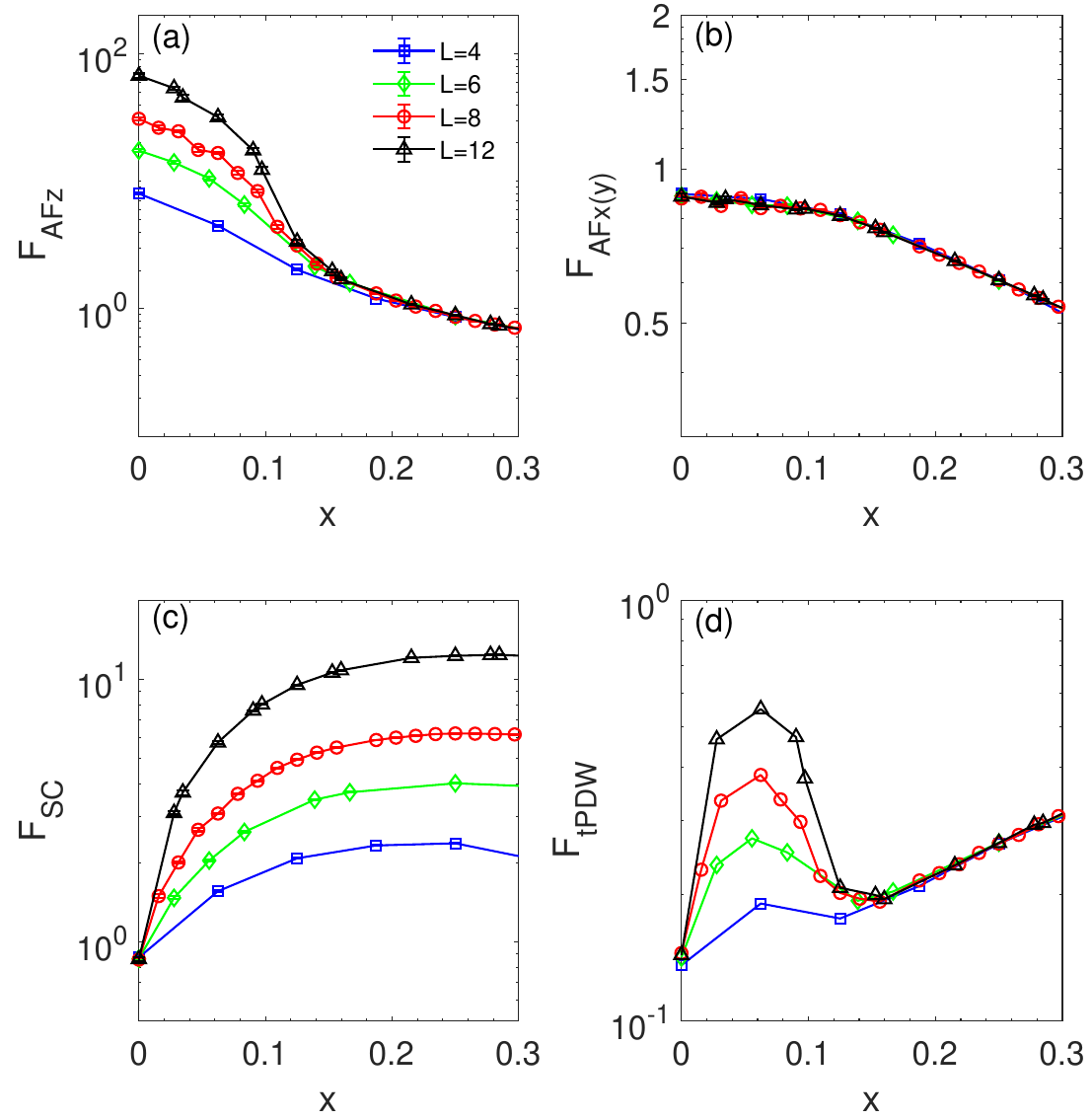}
\caption{\label{Fig:structure}
QMC simulation results for the structure factors of ($a$) AF$_z$,
($b$) AF$_{x(y)}$, ($c$) SC, and ($d$) tPDW versus the doping $x$
as varying $L$.
In the disordered phase for each order parameter, each structure factor
shows very small size-dependence when $L\gg \xi$ and converges to a
value proportional to the square of correlation length $\xi^2$, while in
the ordered phase, it grows as $L\rightarrow\infty$.
The interacting parameter values are $ U=\frac{5}{2},  J_\perp=1, J_z=8,
V=t_\perp=0$.
}
\end{figure}

\noindent
\emph{QMC Results}
We have performed QMC calculations on $2\times L\times L$ lattices with
$L$ up to $12$.
A larger size with $L>12$ is technically difficult  because of the complicated
matrix structures for the general interaction parameters,
which significantly reduces the efficiency of the fast update
algorithm\cite{Assaad2008}.
Even though, our results show clearly a transition from the half-filled
AF insulating phase to the singlet SC phase upon doping.

We first present the QMC simulation results of the structure factors,
defined as the equal-time correlations,
$F(\mathcal{O})=L^2\langle \mathcal{O}^\dag \mathcal{O} \rangle$
where $\mathcal{O}$ represents a physical observable.
In the magnetic channels, $\mathcal{O}$ is chosen as
\bea
N_z&=&\frac{1}{L^2} \sum_i n^4_i(-1)^i, ~~
N_{x(y)}=\frac{1}{L^2}\sum_i n^{2 (3)}_i(-1)^i, \ \ \
\eea
for the AF order along the $z$-direction (AF$_z$),
and that along the $x(y)$-direction (AF$_{x(y)}$),
respectively.
Their structure factors $F_{AF_z}$ and $F_{AF_{x(y)}}$
are shown in Fig.~\ref{Fig:structure} ($a$) and ($b$), respectively.
The structure factors of AF$_z$ increase significantly versus $L$ at
$x<x_c\approx0.11$ indicating the tendency for ordering.
In contrast, those of AF$_{x(y)}$ nearly exhibit no size-dependence,
showing the absence of long-range order.
For the superconducting channel, we have examined
the extended $s$-wave singlet pairing order defined as
$\Delta(i)=\frac{1}{\sqrt{2}L^2}\sum_i(c_{i\uparrow} d_{i\downarrow}-
c_{i\downarrow} d_{i\uparrow} )$,
{\it i.e.}, the pairing across each rung.
Its structure factor $F_{SC}$ increases with enlarging the sample size as
shown in Fig.~\ref{Fig:structure} ($c$).
If expressed with the bonding and anti-bonding band operators,
$f^{e(o)}_\alpha(i) =\frac{1} {\sqrt 2} (c_\alpha(i) \pm d_\alpha(i))$,
this pairing order parameter exhibits opposite signs on the $f^{e,o}$-bases
as
\bea
\Delta(i)=\frac{1}{\sqrt 2}(f^e_\uparrow(i) f^e_\downarrow(i)
-f^o_\uparrow(i) f^o_\downarrow(i)),
\eea
hence, it is an extended $s$-wave
pairing order parameter.
{
We have also measured the superconducting correlations within the layers, but it is much (several orders) smaller than the inter-layer one.
Hence, in the following, only the pairing across each rung will be
considered.
The extended $s$-wave pairing symmetry is among the promising
candidates for the iron-based superconductors \cite{Chubukov_ARCMP_2012}.}
In addition, a triplet pair-density wave (tPDW) correlation is found,
whose order parameter is defined as
\bea
O_{tPDW}=\frac{1}{\sqrt{2}L^2}\sum_i(c_{i\uparrow} d_{i\downarrow}
+ c_{i\downarrow} d_{i\uparrow} )(-1)^i.
\eea
It tends to develop ordering at $0<x<x_c$ even though
its magnitudes are small, as shown in Fig.~\ref{Fig:structure} ($d$).

\begin{figure}
\includegraphics[width=0.45\textwidth]{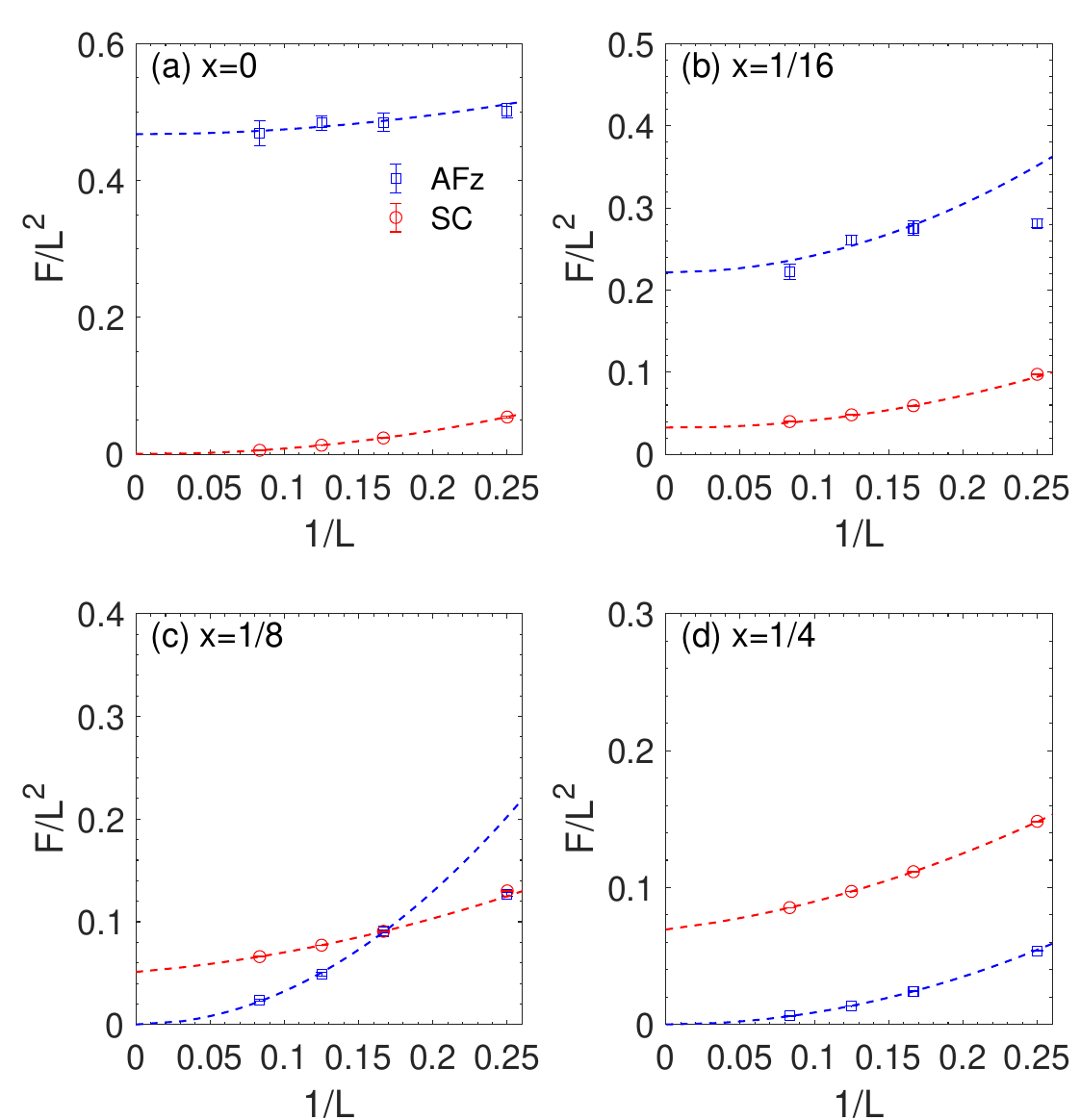}
\caption{The structure factors $F/L^2$ vs $1/L$ for both
the AF$_z$ and SC orders.
They are plotted at doping levels of ($a$) $x=0$,
($b$) $x=\frac{1}{16}$, ($c$) $x=\frac{1}{8}$,
and ($d$) $x=\frac{1}{4}$, respectively.
The dashed lines are polynomial fittings (see the main text for details)
to the QMC data from $L=6$ to $L=12$.
The interaction parameters are the same as in Fig.~\ref{Fig:structure}.
}
\label{Fig:extrapolation}
\end{figure}

{
Next we perform the finite-size scaling for these structure factors to
extract the values of orderings in the thermodynamic limit as
shown in Fig.~\ref{Fig:extrapolation}.
It is based on the scaling hypothesis $F(L)/L^2=a+b/L+c\xi^2/L^2$,
\footnote{{
By assuming the spatial correlation $\langle O(\mathbf{r})O(\mathbf{0})\rangle=a+\frac{b}{4\pi r}+\frac{c}{8\pi}\mathrm{e}^{-r/\xi}$, its Fourier transformation gives the scaling behavior of the structure factor shown in the main text.
Such a scaling method and its generalizations to general polynomials are widely used in literature. \cite{Sorella2012,Assaad2013,Ma2018}}}
where $a$ is the thermodynamic expectation value (square of the order parameter), the $b$-term is the contribution from the gapless excitations (e.g. the Goldstone modes), and the $c$-term represents the short-range correlations, or, the gapped excitations.
For the AF$_z$ order, since there is no Goldstone mode, the $b$-term
is not needed.
However, for the superconductivity, all the three terms should be kept
since there is no Anderson-Higgs mechanism to ``eat'' the gapless
phase mode in our calculations.}

\begin{figure}
\includegraphics[width=0.9\linewidth]{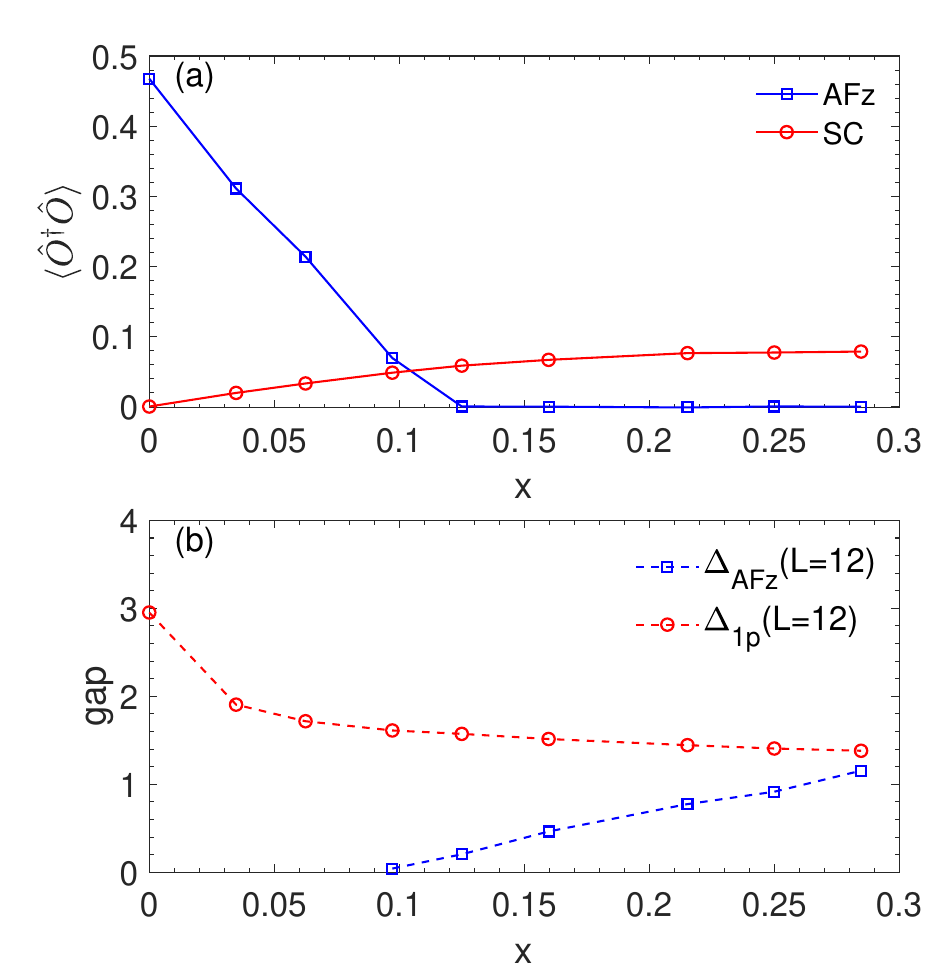}
\caption{\label{Fig:phasediagram}
($a$) The extrapolation of $F/L^2=\avg{O^\dagger O}$ in the limit of $L\to \infty$
versus $x$, where $O$ represents operators for the AF$_z$, and SC order parameters.
The AF$z$ ordering is suppressed beyond a critical doping $x_c\approx 0.11$,
and the SC order coexists with the AF$_z$ one at small dopings $0<x<x_c$.
The interaction parameters are the same as in Fig.~\ref{Fig:structure}.
($b$) The single particle gap $\Delta_{1p}$ at all doping levels and
the spin gap $\Delta_{AF_z}$ at $x>x_c$.
}
\end{figure}

At half-filling ($x=0$), only the AF$_z$ order exhibits a long-range
ordering, while the SC order extrapolates to zero.
At a small doping level with $x=\frac{1}{16}$, the AF$_z$ order still
survives but its value is suppressed accompanied by the emerging of
the SC order.
As the doping level $x$ increases to $\frac{1}{8}$ and above,
the AF$_z$ order vanishes, leaving a pure rung-singlet SC order.
{
The existences of the SC and AF$_z$ long-range orders are also
evidenced by checking the decay patterns of
the two-point correlation functions in real space as shown in the S.M. IV.
In fact, their coexistence is widely seen experimentally such as in
various heavy fermion systems \cite{heavy_fermion_2006,heavy_fermion_2015}}.
All the above results are summarized in the phase diagram as shown in
Fig.~\ref{Fig:phasediagram}.
The AFz order exists in the region of $0<x<x_c$ with $x_c\approx 0.11$,
and the SC order appears immediately upon doping starting from zero.

An interesting observation is that the tPDW tends to develop within
$0<x<x_c$ where the AF$_z$ and SC orders coexist.
Based on the symmetry principle, there exists a coupling among
these three orders constructed as
$L_{tpdw}= g(N_z\Delta^\dag O_{tPDW}+h.c.)$,
where $g$ is an effective coupling constant.
In the coexistence regime, where both $N_z$ and $\Delta$ are finite,
they combine as an external field to induce the tPDW order
although its magnitude is too weak for an accurate identification.
Similarly to the SO(5) theory \cite{Demler2004}, the transition from
the AFz ordering state to the SC state can be unified by
a hidden SO(3) algebra structure:
the total particle number $N$, $O_{tPDW}$ and $O^\dagger_{tPDW}$ form
the generators of a pseudo-spin SO(3) group.
Recently, the pair-density wave, either static or fluctuating,
has received considerably attention due to its potential relation to the CDW and nematic orders in the pseudogap region of high $T_c$ cuprates
\cite{Lee2014,Fradkin2015,Keimer2015,Lu2014,Chen2019}.

We next study the excitation gaps by calculating the imaginary-time-displaced
correlation functions
\bea
\chi(\tau)=\langle T_\tau O(\tau)
O^\dag(0) \rangle,
\eea
where $T_\tau$ means time ordering.
The long-time behavior of $\chi(\tau)$ is related to the excitation gap
$\Delta_O$.
As explained in S. M. V, we measure the mean gap defined as $\Delta_{O}=(E_{O}+E_{O^\dag}-2E_0)/2$
where $E_0$ is the ground state energy and
$E_{O}$ ($E_{O^\dag}$) gives the lowest energy
excited by $O$($O^\dag$).
This gap can be extracted from $\chi(\tau)\chi(-\tau)\sim\mathrm{e}^{-2\Delta_{O}\tau}$ for $\tau\rightarrow\infty$.
For the single-particle gap, $O$ is
chosen as $\psi_\alpha$  with $\alpha = 1 \sim 4$, which yields the diagonal
terms of the single-particle Green's function
$G_{\alpha\alpha}(\tau)$.
We use the averaged results of $G_{\alpha\alpha}(\tau)$ to yield
the single-particle gap $\Delta_{1p}$ as plotted in
Fig.~\ref{Fig:phasediagram} ($a$).
In the whole phase diagram, the single-particle excitations are all
gapped, and $\Delta_{1p}$ reaches the order of the band width in the
antiferromagnetic order dominated region, indicating the existence
of a Mott gap.
We also calculate the spin gap
$\Delta_{AF_z}$ associated with $O = N_z$ in the spin disordered
region, which is also plotted in Fig.~\ref{Fig:phasediagram} ($b$).
It grows up at
$x > x_c$ consistent with the vanishing of the AF$_z$ order.

\begin{figure}
\includegraphics[width=1.0\linewidth]{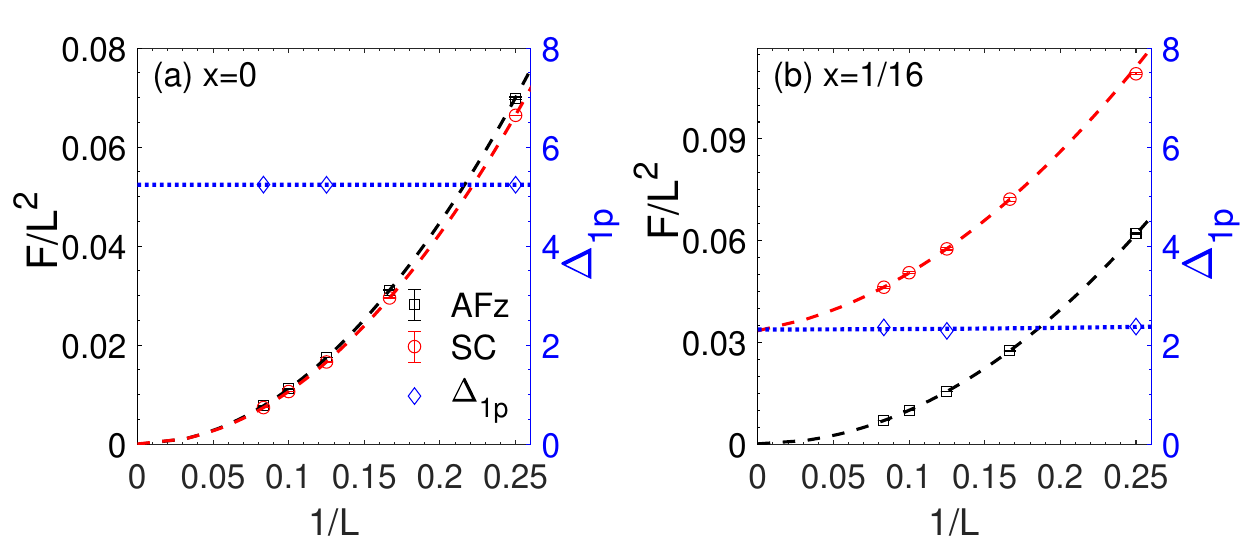}
\caption{\label{Fig:su2extrapolation}
QMC simulations for the SU(2) symmetric model.
The finite size scalings of the structure factors $F/L^2$
of the AF and SC orders, and the single-particle gap $\Delta_{1p}$,
for both at half-filling ($a$) and the $1/16$ doping ($b$).
The scales for $\Delta_{1p}$ are along the axes on the right side
of ($a$) and ($b$).
The interacting parameter values are
$U=4, J_\perp=J_z=\frac{16}{3}$, and $V=t_\perp=0$.
}
\end{figure}

\noindent
\emph{The SU(2) symmetric case}~~
We briefly discuss the consequence if the SU(2) symmetry is preserved.
The QMC simulations are performed by setting $g_{2,3,4}=\frac{16}{3}$
and also $g_c=g_{1,5}=0$,
which corresponds to the case of
$U=4, J_\perp=J_z=\frac{16}{3},  V=t_\perp=0$.
The finite-size scalings of the AF and SC structure factors as well as the single-particle gap $\Delta_{1p}$ at
half-filling and at $x=\frac{1}{16}$ are presented in
Fig.~\ref{Fig:su2extrapolation} ($a$) and ($b$), respectively.
The ground state at half-filling is a Mott insulator
as shown in the nonzero single-particle gap $\Delta_{1p}$ and the
vanishing AF ordering extrapolated to the thermodynamic limit.
Quantum fluctuations are stronger in the SU(2) case than in the
previously studied Ising anisotropic one, hence, the system
is a valence-bond-solid phase without symmetry breaking, i.e.,
the rung-singlet state.
After doping, the SC long-range order is established
in the absence of the AF order,
as shown in Fig.~\ref{Fig:su2extrapolation}($b$),
which is the same as the Ising case.

\noindent
\emph{Summary}~~
In summary, we have performed the projector QMC simulation based
on the auxiliary field method on the bilayer SZH
model, which is free of the sign-problem.
A quantum phase transition occurs from an Ising anisotropic
AF insulating phase, or, an SU(2) invariant
Mott insulating phase without the AF ordering, to
a rung-singlet SC phase with an extended $s$-wave
symmetry driven by doping.
In the coexistence regime between the AF$_z$ and SC
orders, their coupling leads to an enhanced tPDW correlation
as a consequence of the symmetry principle.
This work provides a reliable reference point for studying
superconductivity and other competing orders by doping Mott insulators. Furthermore, the present study can be generalized to other bilayer geometries such as honeycomb or triangular lattices which may be relevant to
certain materials and are left as future works.
%

\noindent
\emph{Acknowledgment}
C. W. thanks Y. Wang for providing computation resource in the early
stage of simulations.
T. M. is supported by Natural Science Foundation of China (NSFC)
(11974049 and 11774033).
D. W. acknowledges the support from NSFC (11874205).
The numerical simulations in this work were performed on HSCC of
Beijing Normal University and Tianhe-II in Guangzhou.

\bibliography{reference}


\newpage

\setcounter{equation}{0}
\setcounter{figure}{0}
\renewcommand{\theequation}{S\arabic{equation}}
\renewcommand{\thefigure}{S\arabic{figure}}
\renewcommand{\thesubsection}{S\arabic{subsection}}

\def\avg#1{\langle#1\rangle}
\def\Re{\rm{Re}}
\def\Im{\rm{Im}}
\def\be{\begin{equation}} \def\ee{\end{equation}}
\def\bea{\begin{eqnarray}} \def\eea{\end{eqnarray}}
\def\PRB{Phys. Rev. B}
\def\PRA{Phys. Rev. A}
\def\PRL{Phys. Rev. Lett.}
\def\nn{\nonumber}
\def\pp{\parallel}

\section*{Supplemental Materials}

\title{Supplemental Materials for ``Doping-driven Antiferromagnetic Insulator - Superconductor
Transition: a Quantum Monte-Carlo Study"}
\author{Tianxing Ma}
\affiliation{Department of Physics, Beijing Normal University,
Beijing 100875, China}
\affiliation{Department of Physics, University of California,
San Diego, California 92093, USA}
\author{Da Wang}
\affiliation{National Laboratory of Solid State Microstructures $\&$
School of Physics, Nanjing University, Nanjing 210093, China}
\author{Congjun Wu}
\affiliation{Department of Physics, University of California,
San Diego, California 92093, USA}

We present the detailed information about
the model Hamiltonian and the quantum Monte Carlo (QMC) method,
including the definition of $\Gamma$-matrices, the projector QMC algorithm,
the scalings of $\Delta \tau$ and $\beta$, the
calculation of excitation gaps, and the spatial correlations.

\subsection{I. Definition of $\Gamma$-matrices}
\label{sect:gamma}
Following the convention in Ref. [\onlinecite{Wu2003}], we define the five
$\Gamma$-matrices as follows:
\bea
\Gamma^1&=&\left (
\begin{array} {cc}
0 & -i I\\
i I& 0
\end{array} \right) , \ \ \
\Gamma^{2,3,4}=\left ( \begin{array}{cc}
{\vec \sigma}& 0\\
0& {-\vec \sigma} \end{array}\right),
\nonumber \\
\Gamma^5&=&\left( \begin{array} {cc}
0& I \\
I & 0 \end{array} \right ),
\eea
where $I$ and
$\vec{\sigma}$ are the 2$\times$ 2 unit and
Pauli matrices.
They satisfy the anti-commutation relation of
\bea
\{ \Gamma^a, \Gamma^b\}=2 \delta_{ab}.
\eea
Their commutators give rise to the 10 generators of the
Sp(4) group as
\bea
\Gamma^{ab}=
-\frac{i}{2} [ \Gamma^a, \Gamma^b] \ \ \ (1\le a,b\le5).
\eea

The identity matrix, $\Gamma^a ~(1\le a \le 5)$ and $\Gamma^{ab} ~(1\le a<b
\le 5)$ span the complete basis for the 16 bilinear operators in the
particle-hole channel for 4-component fermions defined as
\bea
n_i &=& \psi^\dagger_{i,\alpha} \psi_{i,\alpha}, \nonumber \\
n^a_i&=& \frac{1}{2}
\psi^\dagger_{i,\alpha} \Gamma^a_{\alpha\beta} \psi_{i,\beta}, \nonumber \\
L^{ab}_i &=&  -\frac{1}{2}\psi^\dagger_{i,\alpha}
\Gamma^{ab}_{\alpha\beta} \psi_{i,\beta}.
\label{ch3:phbilinear}
\eea

In the context of the bilayer model in the main text, we have
\bea
n_i &=&c^\dagger_{i\sigma} c_{i\sigma} +d^\dagger_{i\sigma} d_{i\sigma},
\nn \\
n^1_i &=&-\frac{i}{2}(d^\dagger_{i\sigma} c_{i\sigma}-h.c.), \nn \\
n^5_i &=&  \frac{1}{2}(d^\dagger_{i\sigma} c_{i\sigma}+h.c.),  \nn \\
n^{2,3,4}_i &=& c^\dagger_{i,\alpha}
\left(\frac{\vec \sigma}{2}\right)_{\alpha\beta}c_{i\beta}
-d^\dagger_{i,\alpha}
\left(\frac{\vec \sigma}{2}\right)_{\alpha\beta}d_{i\beta},
\eea
where $n_i$ is the total particle number on the rung,
$n^{1}_i$ and $n^{5}_i$ are the bond current and bond strength
along the rung, respectively, and $n^{2,3,4}$ are the bond N\'eel order.
We define the Kramers symmetry as
\bea
{\cal T}=\Gamma^1\Gamma^3 C,
\eea
where $C$ is the complex conjugate.
Physically, ${\cal T}$ is the combination of the usual time-reversal
transformation and the flipping of the upper and lower layers.
It is easy to check that the above 6 bilinear operators are even
under this Kramers operations.

The other 10 bilinear operators are odd under ${\cal T}$, which can be
organized as
\bea
\mbox{Re} \vec \pi_i &=& c^\dagger_{i\alpha} \left( \frac{\vec \sigma}{2}
\right)_{\alpha\beta} d_{i\beta} +h.c. ,\nn \\
\mbox{Im} \vec \pi_i &=& -i \big [c^\dagger_{i\alpha} \left( \frac{\vec \sigma}{2}
\right)_{\alpha\beta} d_{i\beta} -h.c. \big], \nn \\
\vec S_i &=& c^\dagger_{i,\alpha}
\left(\frac{\vec \sigma}{2}\right)_{\alpha\beta}c_{i\beta}
+d^\dagger_{i,\alpha}
\left(\frac{\vec \sigma}{2}\right)_{\alpha\beta}d_{i\beta},
\nn \\
Q_i &=&\frac{1}{2}(c^\dagger_{i\sigma} c_{i\sigma}
-d^\dagger_{i\sigma} d_{i\sigma}),
\eea
where $\mbox{Re}\vec \pi_i $ is the spin-channel bonding strength,
$\mbox{Im}\vec \pi_i $ is the spin current along the rung,
$\vec S_i $ is the total spin of the rung,
and $Q_i$ is the charge-density-wave order of the rung.

\subsection{II. The projector QMC algorithm}
\label{sect:QMC}

We adopt the projector determinant QMC method \cite{Assaad2008} to study
the model Hamiltonian shown in Eq. 1 in the main text.
The basic idea is to apply
the  projection operator $\mathrm{e}^{-\beta H/2}$
on a trial wave function $|\Psi_T\rangle$.
If $\avg{\Psi_G|\Psi_T}\neq 0$ and there exists
a nonzero gap between $|\Psi_G\rangle$
and the first excited state, $|\Psi_G\rangle$ is arrived
as the projection time $\beta\rightarrow \infty$,
\begin{equation}
|\Psi_G\rangle=\lim_{\beta\rightarrow\infty}\mathrm{e}^{-\beta H/2}|\Psi_T\rangle,
\end{equation}
where the projection time $\beta$ can be divided into $M$ slices
with $\beta=M\Delta\tau$, and the trial wave function can be written by filling $N_e$ electrons,
\begin{equation}
|\Psi_T\rangle=\prod_{i}\sum_{j=1}^{N_e}c_j^{\dag} P_{ji} |0\rangle.
\end{equation}
Here $i,j$ contains both site and flavor indices and $|0\rangle$ labels
the fermion vacuum.
In practice, $|\Psi_T\rangle$ can be chosen as the ground state of a free fermion Hamiltonian.
The scattering matrix $\langle \Psi_T|\mathrm{e}^{-\beta H}|\Psi_T\rangle$ is obtained by integrating out the fermionic degrees of freedom,
\bea
\langle \Psi_T|\mathrm{e}^{-\beta H}|\Psi_T\rangle
=\sum_{\{\sigma\}}[\prod_{i}\gamma_i(\sigma_i)]
\mathrm{det}(P^{\dag}B_LB_{L-1}...B_1P),
\nonumber \\
\label{eq:Smatrix}
\eea
where $\sigma_i$ labels the auxiliary discrete boson field (see below).
The scattering matrix Eq.~\ref{eq:Smatrix}, which plays the role of the partition function, serves as the basis of the projector determinant QMC algorithm. The $\{\sigma_i\}$ fields are then sampled by using
the standard Monte Carlo technique.

In order to obtain Eq.~\ref{eq:Smatrix}, two preliminary steps are needed.
The second order Suzuki-Trotter decomposition
\bea\label{eq:trotter}
\mathrm{e}^{-\Delta\tau(K+V)}=\mathrm{e}^{-\Delta\tau K/2}
\mathrm{e}^{-\Delta\tau V}\mathrm{e}^{-\Delta\tau K/2}+o[(\Delta\tau)^3]
\eea
is first used to separate
the kinetic ($K$) and interaction ($V$) terms in each time slice, and
then the $\mathrm{e}^{-\Delta\tau V}$ term is decoupled by using
the discrete Hubbard-Stratonovich  transformation,
\bea
e^{gX^2}= \sum_{\sigma=\pm1,...\pm I_{\rm max}}\gamma(\sigma)
e^{\lambda(\sigma)X},
\label{eq:approxHS}
\eea
where $\sigma$ is the discrete Hubbard-Stratonovich field.
If eigenvalues $\mathrm{eig}(X)=\{0,\pm1\}$, the maximal value of
$\sigma$, $I_{max}$ can be set as 1 \cite{Hirsch1983} along with
the choices of $\gamma(\sigma)$ and $\lambda(\sigma)$ as
\bea
\gamma(\pm1)=\frac{1}{2},~~ \lambda(\pm1)= \pm \cosh^{-1}(e^g).
\eea
If $\mathrm{eig}(X)=\{0,\pm1,\pm2 ,\pm3\}$, we need set $I_{\rm max}=2$
and choose
\bea
\gamma(\pm 1)&=&\frac{-a(3+a^2)+d}{4d}, \nn\\
\gamma(\pm 2)&=&\frac{a(3+a^2)+d}{4d}, \nn \\
\eta(\pm 1)&=&\pm\cosh^{-1} \left\{ \frac{a+2a^3+a^5+(a^2-1)d}{4}
\right\} \nn\\
\eta(\pm 2)&=&\pm\cosh^{-1} \left \{\frac{a+2a^3+a^5-(a^2-1)d}{4}
\right\},\nn \\
\label{eq:newHS}
\eea
where $a=e^{g}$, $d=\sqrt{8+a^2(3+a^2)^2}$ \cite{Wang2014}.
In our case, $X=\psi^\dag \Gamma^{2,3,4}\psi$, whose eigenvalues
are among $0,\pm 1, \pm 2$, hence, the latter Hubbard-Stratonovich
transformation is applied.

\begin{figure}
\includegraphics[scale=0.4]{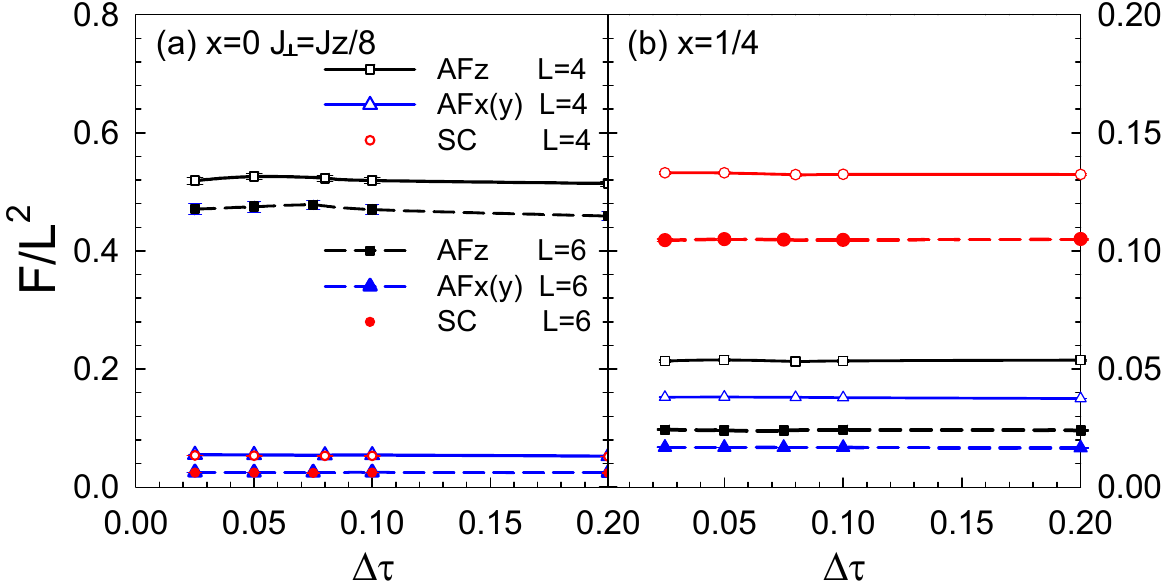}
\centering
\caption{(Color online) The $\Delta\tau$-dependence of the structure
factors for various order parameters for $x=0$ ($a$) and
$x=\frac{1}{4}$ ($b$) with $L=4$ (solid lines) and $L=6$ (dashed lines).
AFz, and AFx(y) represent the antiferromagnetic order along the $z$-direction,
and that along the $x$ or $y$-direction, respectively,
and SC represents the superconducting order.
The interacting parameter values are $U=2.5, V=t_\perp=0, J_\perp=1$,
and $J_z=8$.
}
\label{Fig:checkdt}
\end{figure}

\begin{figure}
\includegraphics[scale=0.4]{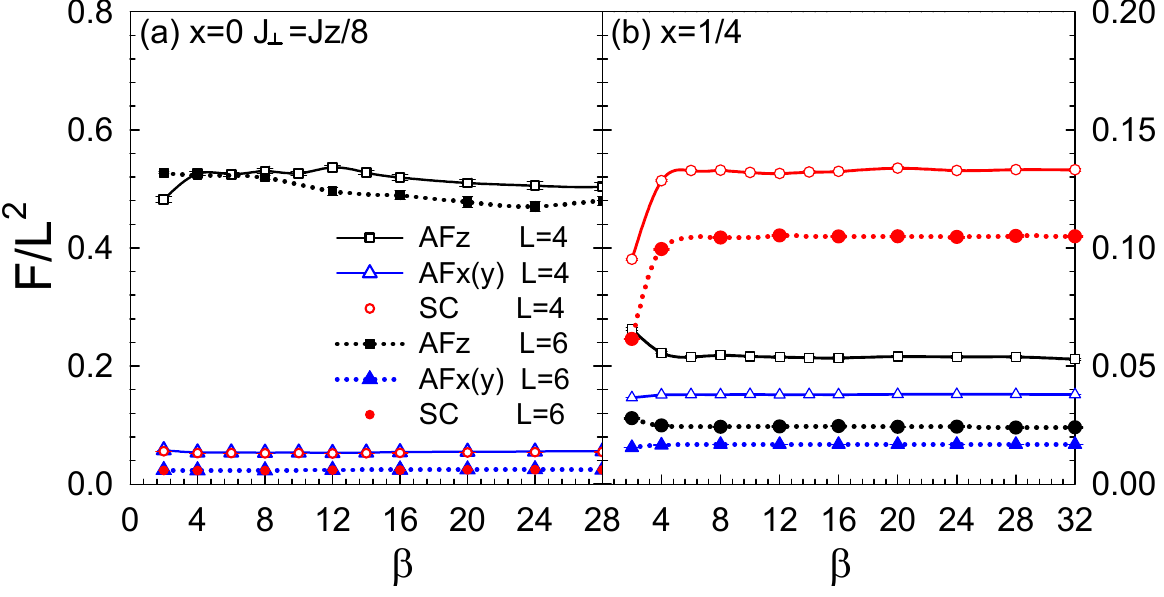}
\centering
\caption{(Color online) The $\beta$-dependence of structure factors of
various order parameters for $x=0$ ($a$) and $x=\frac{1}{4}$ ($b$)
with $L=4$ (solid lines) and $L=6$ (dashed lines).
The symbols and the interacting parameter values are the same
as those presented in Fig. \ref{Fig:checkdt}.
}
\label{Fig:checkbeta}
\end{figure}

\subsection{III. The $\Delta \tau$ and $\beta$-scalings}
\label{sect:errorB}

In the projector QMC algorithm, the systematic error mainly comes from
two origins: the finite time step $\Delta\tau$ and the finite projection
time $\beta$.
In the following, we perform the error analysis on
both $\Delta\tau$ and $\beta$.
In this section, we employ the parameter values
for $U=2.5, J_\perp=1, J_z=8$, $V=t_\perp=0$ for
simulations below.

For the Suzuki-Trotter decomposition defined in Eq.~\ref{eq:trotter},
detailed calculation shows that its error is at the order of ${\rm max}\{tg_i^2,t^2g_i\}(\Delta\tau)^3$.
In Fig. \ref{Fig:checkdt}, scalings of the antiferromagnetic structure
factors along the $z$, $x(y)$ directions, and the
superconductivity structure factor {\it v.s.} $\Delta\tau$ are plotted
for $x=0$ in ($a$) and $x=\frac{1}{4}$ in ($b$).
The slopes of these scaling lines are nearly independent on the lattice
size $L$ for all three orders.
Therefore, we only need to check the small lattice size.
Due to the convergence of the finite $\Delta\tau$ scaling, we
use the value of $\Delta\tau=0.1$ in all the simulations.

We further check the effect of the finite projection time $\beta$.
In Fig.~\ref{Fig:checkbeta}, the scalings of the antiferromagnetic
structure factors along the $z$ and $x(y)$ directions, and the
superconducting structure factor {\it v.s.} $\beta$ are presented.
For each curve, $\beta_c$ is defined as the convergence projection time
after which the structural factors converge.
It is shown that the antiferromagnetic order parameter along the $x (y)$
direction and the superconducting order parameter converge very quickly
for both $x=0$ and $x=1/4$.
The corresponding $\beta_c$ is found to be around $8$.
For the antiferromagnetic order along the $z$-direction, we set $\beta_c=16$
should be enough for $L=4$, and $\beta_c=24$ for $L=6$ as well.
This indicates that $\beta_c(L) = 4L$ is safe for convergence,
which is taken for all the simulations presented in the main text
for accurate numeric results.

\subsection{IV. Spatial correlations}

\begin{figure}[h]
\includegraphics[width=0.45\textwidth]{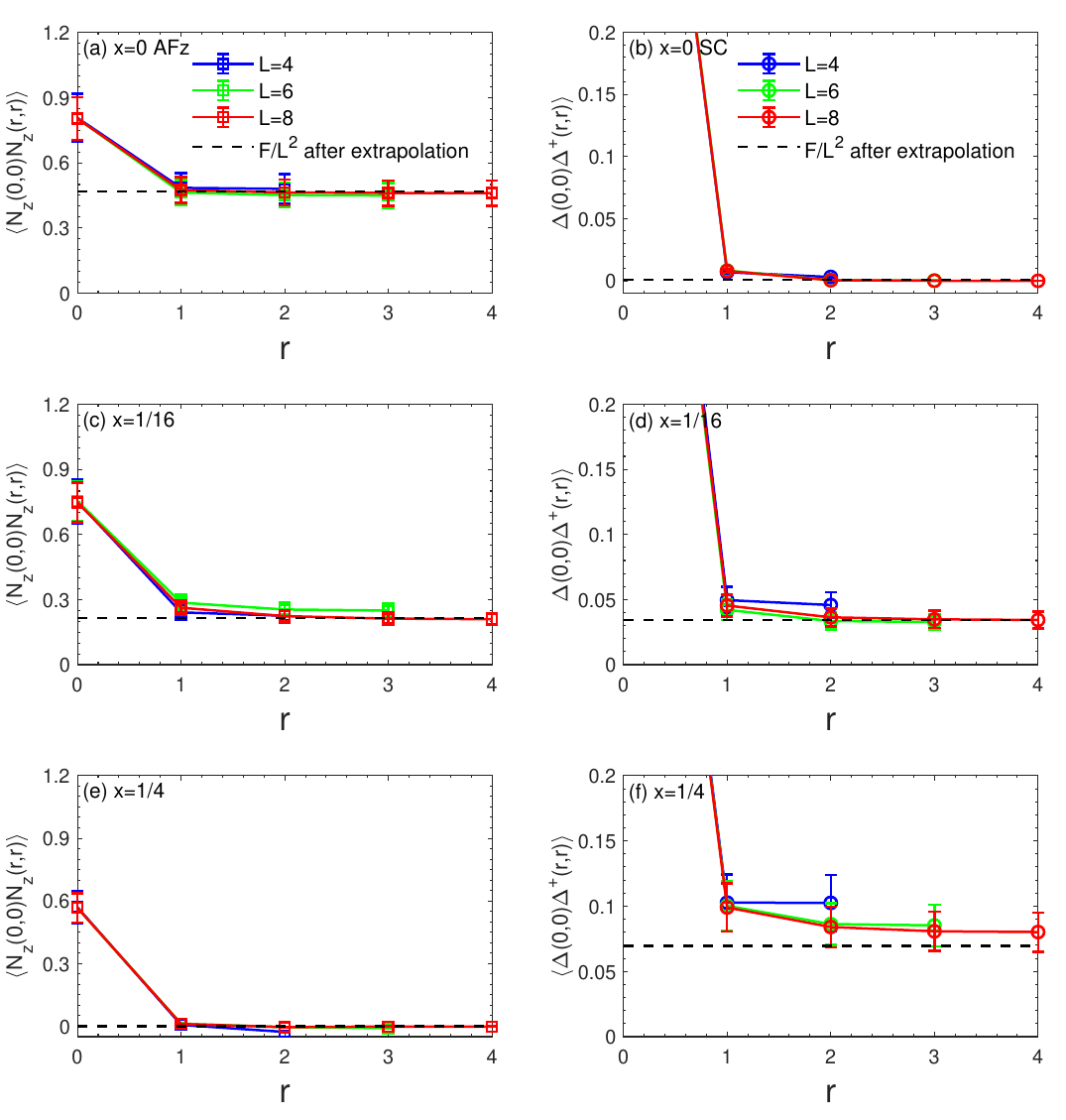}
\caption{\label{Fig:spatial}
The spatial correlation functions at different doping levels
with system size $L=4,6$ and $8$.
The correlation functions for the AFz and SC orders at $x=0$ are
plotted in ($a$) and ($b$), respectively;
those at $x=1/16$ are plotted in ($c$) and ($d$) ( $x=1/18$ for the case of
$L=6$), respectively;
those at $x=1/4$ are plotted in  ($e$) and ($f$), respectively.
The squares of the order parameters obtained by the finite size scaling
on the structure factors in the main text are plotted with dashed lines
for comparison.
}
\end{figure}
{
To demonstrate the SC long-range order after doping and its coexistence with the AFz order, we examine their spatial correlations
$\langle O(0,0)O^\dag(r,r)\rangle$ \cite{Clay2008}.
The results are presented in Fig. 4 at three typical doping levels, i.e., $x=0, 1/16$, and $1/4$, which correspond to the cases with only the AFz order, the coexistence of the AFz and SC orders, and only the SC order, respectively. For the system size with $L=6$, $x=1/18$ is used instead due to its commensurability with the system.

For all of these doping levels, the spatial correlations saturate at large distances. As $L$ increases, the farthest correlation functions approach the values obtained via the finite-size scalings on the corresponding structure factors in the main text. The consistency between two approaches demonstrates that the long-rang orderings of the AFz and SC are reliable.}

\subsection{V. Calculation of excitation gaps}

\begin{figure}
\includegraphics[width=0.45\textwidth]{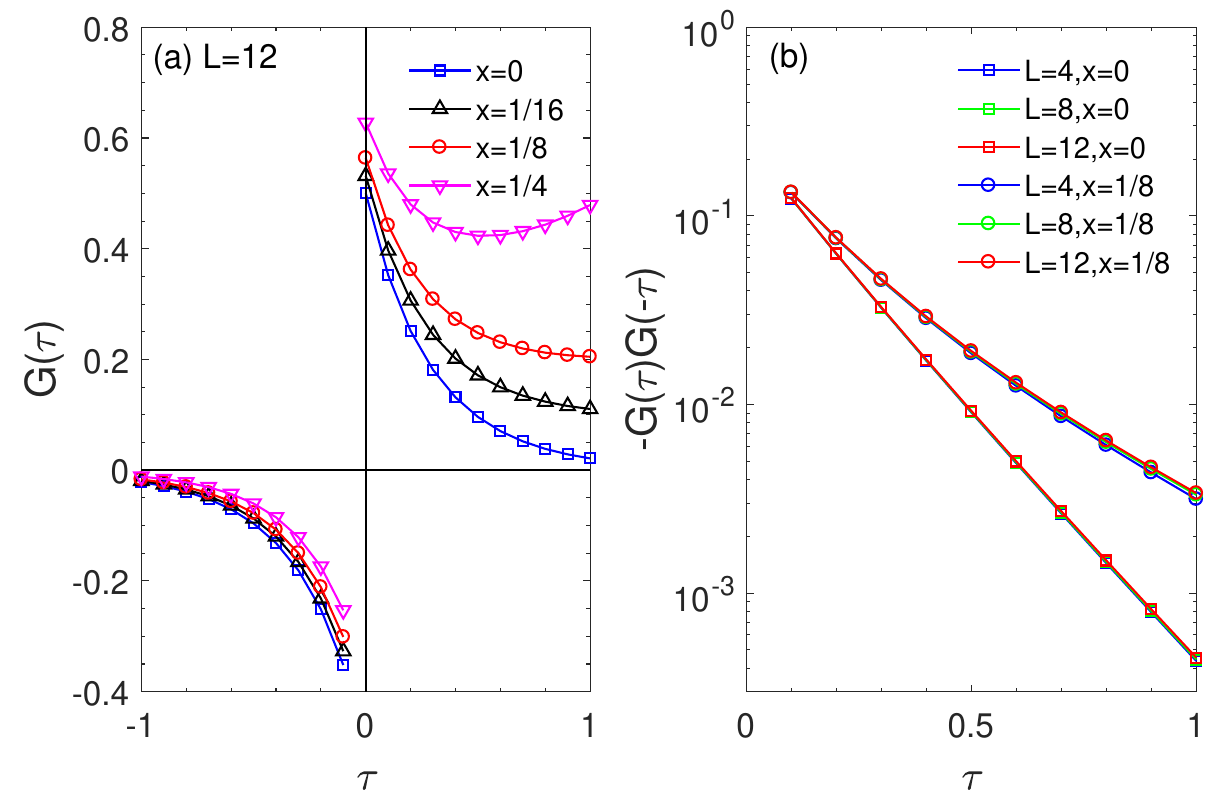}
\caption{\label{Fig:gtau}
The imaginary time Green's functions.
$G(\tau)$ is plotted at different dopings.
Due to the particle-hole symmetry at half-filling, the relation of
$G(\tau)=-G(-\tau)$ is satisfied, while this symmetry is not
held away from half-filling.
We employ $[-G(\tau)G(-\tau)]$ to extract the mean single
particle gap, plotted in (b), which shows very weak
size dependence.}
\end{figure}

As explained in the main text, we calculate the spectra gap functions
through the imaginary-time displaced correlation functions
$\chi(\tau)=T_\tau \avg{ {\cal O}(\tau) {\cal O^\dagger}(0)}$.
Since our QMC works in the canonical ensemble, we can only obtain
the energy difference directly through $\chi(\tau)\sim\mathrm{e}^{-(E_{\mathcal{O}^\dagger}-E_0)\tau}$ for $\tau\rightarrow\infty$ and $\chi(\tau)\sim\mathrm{e}^{(E_{\mathcal{O}}-E_0)\tau}$ for $\tau\rightarrow-\infty$, where $E_0$ is the ground state energy and $E_{\mathcal{O}}$($E_{\mathcal{O}^\dag}$) gives the lowest energy
excited by $\mathcal{O}$($\mathcal{O}^\dag$).
On the other hand, the physical gap should take the chemical potential
into account, i.e. $\Delta_{\mathcal{O}}=E_{\mathcal{O}}-E_0-
\mu N_{\mathcal{O}}$ where $N_\mathcal{O}$ is the particle number
of the excited states.
Nevertheless, the relation between particle number $N$ and $\mu$
is generally complicated especially for an interacting model.
We use the average of $\Delta_{\mathcal O}$ and
$\Delta_{\mathcal{O^\dagger}}$ as the excitation
gap, in which $\mu$ does not appear explicitly.

In Fig.~\ref{Fig:gtau} ($a$), we plot the single-particle Green's
function $G(\tau)$ as an example to clarify our points.
Only at half-filling, $G(\tau)$ shows the particle-hole
symmetry, {\it i.e.}, $G(\tau)=-G(-\tau)$.
Away from the half-filling, the particle-hole symmetry is broken.
If we directly take the slope of $\log[G(\tau)]$ versus $\tau$
as the excitation gap, we even obtain a negative value,
for example, at $x=1/4$.
According to the above discussions, we extract the mean gaps from
$\log[-G(\tau)G(-\tau)]$, as shown in Fig.~\ref{Fig:gtau}($b$),
which show very small size-dependences.


%
%
%
%
%
%
%

\end{document}